\documentclass[aip,rsi,graphicx,amsmath,amssymb,reprint]{revtex4-1}
\usepackage{t1enc}
\usepackage[latin1]{inputenc}
\usepackage{graphics}
\usepackage{graphicx}
\usepackage{amsmath,amsthm,amsfonts,amssymb,amsxtra,amsopn}
\usepackage{color}
\usepackage{sistyle}
\usepackage{exscale} % f"ur die Gr"o"senanpassung von /int und /sum %
\usepackage{epsfig} % Skalierung von eps-Bildern %

\begin{document}

% Use the \preprint command to place your local institutional report number 
% on the title page in preprint mode.
% Multiple \preprint commands are allowed.
%\preprint{}

\title{An ultrahigh-vacuum cryostat for simultaneous scanning tunneling microscopy and magneto-transport measurements down to 400\,mK}

% repeat the \author .. \affiliation  etc. as needed
% \email, \thanks, \homepage, \altaffiliation all apply to the current author.
% Explanatory text should go in the []'s, 
% actual e-mail address or url should go in the {}'s for \email and \homepage.
% Please use the appropriate macro for the type of information

% \affiliation command applies to all authors since the last \affiliation command. 
% The \affiliation command should follow the other information.

\author{Marcus Liebmann}
\email[]{liebmann@physik.rwth-aachen.de}
%\homepage[]{Your web page}
%\thanks{}
%\altaffiliation{}
\affiliation{II.\ Inst.\ Phys.\ B and JARA-FIT, RWTH Aachen University, 52074 Aachen, Germany}

\author{Jan Raphael Bindel}
\affiliation{II.\ Inst.\ Phys.\ B and JARA-FIT, RWTH Aachen University, 52074 Aachen, Germany}

\author{Mike Pezzotta}
\affiliation{II.\ Inst.\ Phys.\ B and JARA-FIT, RWTH Aachen University, 52074 Aachen, Germany}

\author{Stefan Becker}
\affiliation{II.\ Inst.\ Phys.\ B and JARA-FIT, RWTH Aachen University, 52074 Aachen, Germany}

\author{Florian Muckel}
\affiliation{II.\ Inst.\ Phys.\ B and JARA-FIT, RWTH Aachen University, 52074 Aachen, Germany}

\author{Tjorven Johnsen}
\affiliation{II.\ Inst.\ Phys.\ B and JARA-FIT, RWTH Aachen University, 52074 Aachen, Germany}

\author{Christian Saunus}
\affiliation{II.\ Inst.\ Phys.\ B and JARA-FIT, RWTH Aachen University, 52074 Aachen, Germany}

\author{Christian R.\ Ast}
\affiliation{Max-Planck-Institut f{\"u}r Festk{\"o}rperforschung, 70569 Stuttgart, Germany}

\author{Markus Morgenstern}
\affiliation{II.\ Inst.\ Phys.\ B and JARA-FIT, RWTH Aachen University, 52074 Aachen, Germany}

\date{\today}

\begin{abstract}
We present the design and calibration measurements of a scanning tunneling microscope setup in a $^3$He ultrahigh-vacuum cryostat operating at 400\,mK with a hold time of 10 days. With 2.70\,m in height and 4.70\,m free space needed for assembly, the cryostat fits in a one-story lab building. The microscope features optical access, an $xy$ table, {\it in-situ} tip and sample exchange, and enough contacts to facilitate atomic force microscopy in tuning fork operation and simultaneous magneto-transport measurements on the sample. Hence, it enables scanning tunneling spectroscopy on microstrucured samples which are tuned into preselected transport regimes. A superconducting magnet provides a perpendicular field of up to 14\,T. The vertical noise of the scanning tunneling microscope amounts to $1\,\mathrm{pm_{rms}}$ within a 700\,Hz bandwidth. Tunneling spectroscopy using one superconducting electrode revealed an energy resolution of $120\,\mu$eV. Data on tip-sample Josephson contacts yield an even smaller feature size of $60\,\mu\mathrm{eV}$, implying that the system operates close to the physical noise limit.
\end{abstract}

% insert suggested PACS numbers in braces on next line

\maketitle %\maketitle must follow title, authors, abstract and \pacs

% Body of paper goes here. Use proper sectioning commands. 
% References should be done using the \cite, \ref, and \label commands
\section{Introduction}
\label{intro}
Scanning tunneling microscopy (STM) and spectroscopy (STS) give access to to electronic wavefunctions down to the atomic limit.\cite{crommie93} They allow to probe different excitations such as phonons,\cite{smith86} vibrons,\cite{brune93} magnons,\cite{balashov06} or single spin-flips.\cite{heinrich04} They enable the investigation of atomic-sized contacts including atomic chains\cite{pascual93}, functionalized tips\cite{kelly96} or light emission.\cite{berndt93} They provide time resolution down to 100\,ps\cite{loth10} or, if combined with pulsed lasers, down to 500\,fs.\cite{cocker13} Recently, even the option of electron spin resonance experiments down to the single atom limit has been shown,\cite{choi17} which opens the door towards ultrahigh-vacuum (UHV) based experiments of coherent dynamics on the atomic scale. Hence, the STM starts to bridge the field of surface science with its well-controlled sample preparation down to the atomic limit and the field of magnetotransport studies with its versatile tuning knobs controlling electron systems down to individual electrons.

A central challenge while moving away from the single crystal based UHV preparation is that one requires much more sophisticated sample holders. Only then, one can make full use of the electronic control of a nanostructured sample. It requires, e.g., the option to pre-position the tip with an accuracy of better than $10\,\mu\mathrm{m}$ using optical means. If areas close to an insulator, e.g., the sample edge, are to be investigated, the tip-sample distance has to be controlled independently of a tunneling current, e.g., by a simultaneously operated atomic force microscope (AFM). For transport measurements, one applies a few gate voltages and a source-drain current in order to tune in a certain transport regime. At the same time, it is mandatory to keep the electronic and vibrational noise level as well as the spatial resolution at highest standards. Finally, a sufficiently low temperature is needed for highest spectral resolution as well as for the stability of the probed electronic phase or regime.

There are currently a number of ultralow-temperature systems, including commercially available ones. However, the high performance demands when combining low-temperature physics and scanning probe microscopy still require unique home-built systems, specialized to the individual needs. Latest systems reaching temperatures below 1\,K feature dilution refrigerators,\cite{kambara07,song10,suderow11,assig13,misra13,singh13,roychowdhury14,galvis15} Joule-Thompson refrigerators,\cite{zhang11} pulse tubes,\cite{pelliccione13,haan14} and sorption-pumped $^3$He systems.\cite{pan99,kugler00,wiebe04,kamlapure13,allwoerden16} Only four of them offer optical access to operate a positioning device capable of $\mu$m adjustment,\cite{kugler00,song10,zhang11,misra13} which is necessary, e.g., to locate exfoliated graphene flakes below the tip.\cite{geringer09,mashoff10,freitag16} The optical access in combination with UHV transfer and magnetic field usually results in huge vacuum recipients occupying more than one floor for handling of the low-temperature insert.\cite{kugler00,song10,misra13} This excessive height also poses a difficult environment for vibration isolation so that none of these instruments have reported a vertical noise of less than 2\,pm.

We present a more compact ultrahigh vacuum (UHV) system fitting in a one-floor room. It contains preparation and analysis chambers hosting a home-built scanning tunneling microscope within a sorption-pumped $^3$He cryostat and operates in a perpendicular magnetic field of 14\,T. The microscope features five contacts for {\it in-situ} transport measurements of the sample while simultaneously scanning with a tuning-fork-operated scanning probe tip. It provides optical access and the possibility of micropositioning down to 10\,$\mu$m while maintaining a compact design suitable for a one-story lab space. It moreover provides enough stability for high-resolution scanning tunneling spectroscopy (STS) down to 0.1\,meV. Temperature-dependent measurements can be conducted in the range of $0.4 - 9$\,K.

\section{Design}
\subsection{The UHV system}
The three main sections are the preparation chamber, the analysis chamber and the cryostat. All chambers are connected via an {\it in-situ} transfer system. Samples can be loaded from air into the system through a load lock connected to the preparation chamber. It is also compatible with a home-built vacuum transfer shuttle which is equipped with a mobile getter pump and a wobble stick providing the possibility to exchange samples between various UHV systems of our group or distant collaboration partners while maintaining a pressure of $5 \cdot 10^{-8}$\,Pa. The preparation chamber is designed for sample annealing using a resistance heater and electron beam heater for temperatures up to $T = 3700$\,K, sputtering or plasma-based ion bombardment.\cite{anton00} From the preparation chamber, samples can be transferred into the analysis chamber or straight to the cryostat. The analysis chamber is designed for cross-check type sample investigations such as low energy electron diffraction (LEED) and Auger electron spectroscopy (AES). It also exhibits a port for an additional room temperature STM. Furthermore, it is aided by six ports, one of which can be used for an exchange of a home-built evaporator without breaking the UHV.

\begin{figure*}
\includegraphics{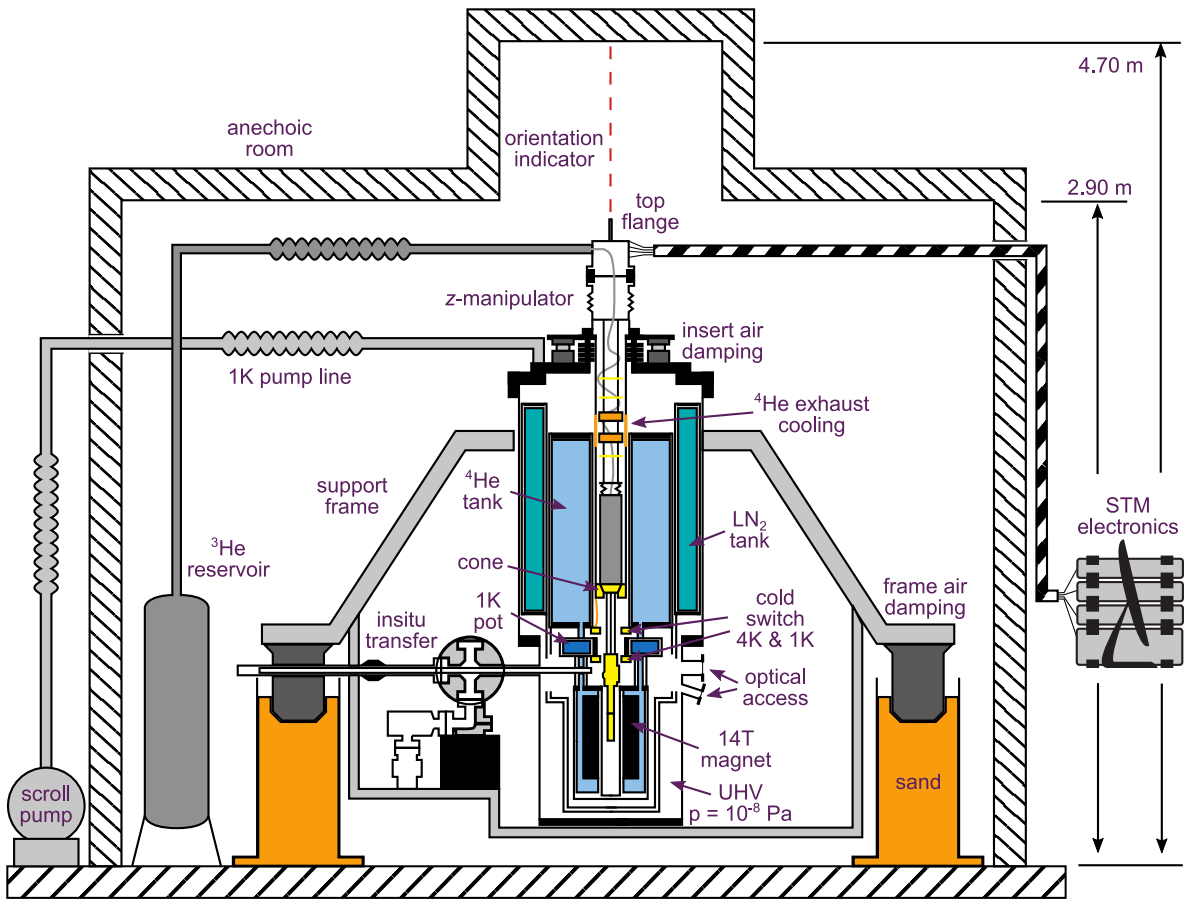}
\caption{\label{fig:L003} Schematic cut through the UHV system. Mechanical pumps operated during measurements and STM electronics are located outside the anechoic room. Air damping isolators placed on sand columns separate the system from the lab floor and the $^3$He insert from the rest of the system. For assembling the cryostat, a free height of 4.70\,m is provided by the dome directly above the cryostat, while the main ceiling height of 2.90\,m is sufficient for normal operation.}
\end{figure*}

A sketch of the UHV system is shown in Figure \ref{fig:L003}.  
The instrument is located on the ground floor in an anechoic room made of bricks and concrete and an acoustic door (damping 55\,dB), and positioned on passive air-pressurized vibration isolators (resonance frequency 1.5\,Hz vert., 2.5\,Hz horiz.). The stainless steel support frame has been optimized by finite element simulations\cite{solidworks} towards low vibration amplitudes and is additionally filled with sand for further damping. Mechanical pumps (turbo-molecular pump and roughing pump) for obtaining UHV conditions by bakeout and during sample preparations are also located in the anechoic room but can be switched off during measurements. A base pressure of $2 \cdot 10^{-8}$\,Pa is maintained by ion getter and Ti sublimation pumps. Electronics and pumps for operating the cryostat as well as the experimenter stay outside the anechoic room.

\subsection{The Cryostat}
The cryostat was designed and built in cooperation with {\it CryoVac}.\cite{cryovac} It is completely built in UHV technology, i.e., the $^4$He tank can be baked to 80\,$^{\circ}$C limited by the magnet, and to 120\,$^{\circ}$C for the other parts if the magnet temperature is carefully controlled. It consists of three main parts: an outer tank for liquid Nitrogen (LN$_2$), an inner tank for liquid $^4$He and a $^3$He insert hosting the microscope. The different temperature levels are shielded by exhaust-cooled radiation shields. The LN$_2$ tank is filled with fine copper wool in order to prevent the liquid Nitrogen from boiling. However, the tank could also be pumped for solidifying the cryogen, if necessary. The $^4$He tank is divided into three parts: the main tank is surrounded by the LN$_2$ tank. Below, there is the magnet tank connected with CF40 flanges and hosting a 14\,T superconducting solenoid magnet. This part is surrounded by LN$_2$ and $^4$He cooled radiation shields. The main tank allows a hold time of six days (similar for LN$_2$), and the magnet can be kept in persistent mode during refilling. Between main tank and magnet tank, the 1\,K pot is located. Its volume of 5\,liters can be filled via a direct line and a cold valve from the main tank and subsequently pumped by a scroll pump (pumping speed: $35\,\mathrm{m}^3\mathrm{s}^{-1}$) outside the anechoic room. This allows to reach a temperature of 1.5\,K. Below the 1\,K pot, the microscope can be positioned for {\it in-situ} tip and sample transfer. Access is given by rotating parts of the shields horizontally using a wire rope and a rotational feedthrough. All tanks and shields are surrounded by UHV. In normal operation, the cryostat fits into a lab with 2.90\,m height. For assembly, a height of 4.70\,m is necessary. All parts necessary fit through a lab door with a height of 2.00\,m. In principle, the assembly of the cryostat outside the lab would also be possible, requiring access doors of 2.50\,m in height and 1.20\,m in width.

\subsubsection{Insert design}
\begin{figure}
\includegraphics{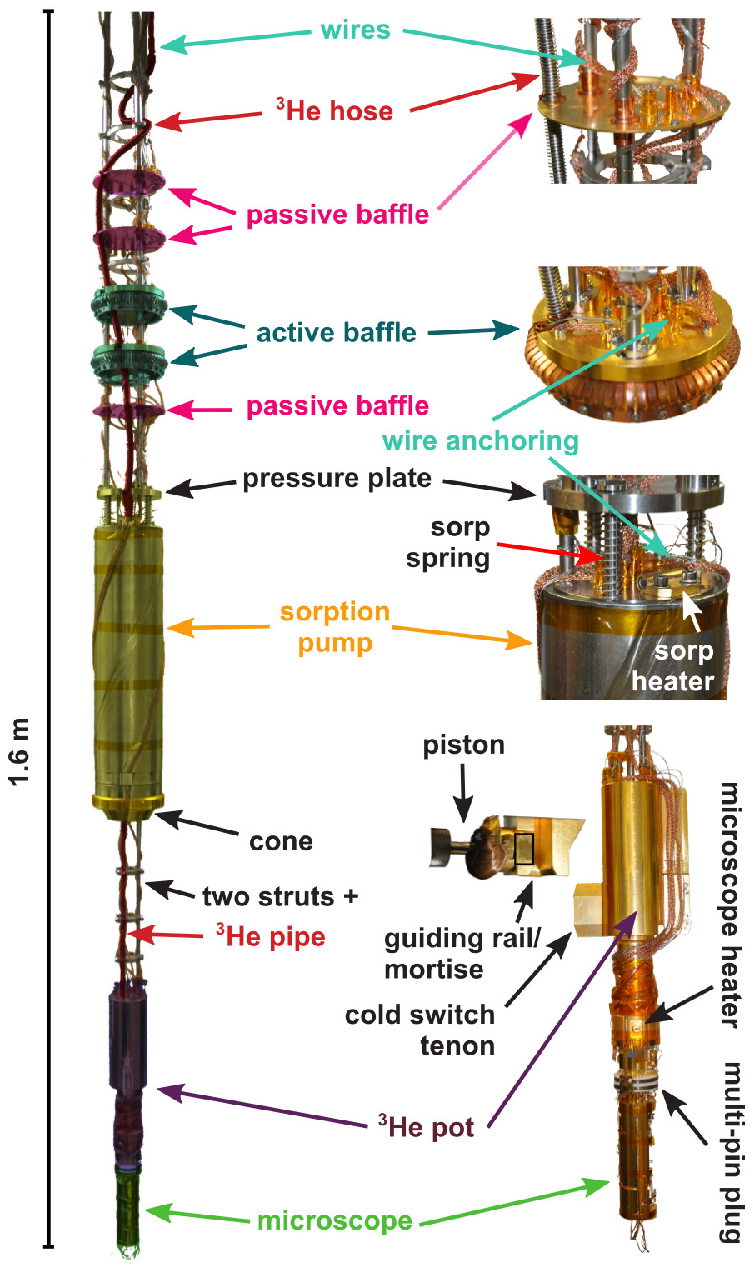}
\caption{\label{fig:insert} Photography of the $^3$He insert (left) with selected zoom-ins (right). The whole assembly hangs down in UHV from electrical feedthrough flanges on top of a $z$ manipulator (not shown) and is surrounded by a pipe carrying the OFHC copper counterparts for the active baffles and the sorption pump cone (also not shown). Wires are anchored at each baffle, at the sorption pump and the $^3$He pot. The piston, guiding rail and mortise exist at two positions realizing two cold switches which clamp the tenon of the $^3$He pot at two different heights of the $z$ manipulator.}
\end{figure}
The last cooling step is conducted with the $^3$He insert which hangs from the top flange of a $z$ manipulator mounted on the top flange of the vacuum recipient. It is mounted on a second set of passive air-pressurized vibration isolators (resonance frequency 3\,Hz vert., 5\,Hz horiz.). Thus, vibrations arising from boiling cryogen, from turbulences of the exhaust gas, or from thermoacoustic oscillations of the $^4$He gas are damped. The $z$ manipulator serves as a tool for moving the microscope between the transfer plane of the UHV system and the measurement position in the center of the magnet. It also enables different parts of the insert to be coupled to different temperature levels necessary during a cooling cycle or regeneration cycle of the sorption pump.

Figure \ref{fig:insert} shows the low-temperature part of the insert. At the bottom, a $^3$He pot made from gold-plated OFHC copper is located. It can host 100\,ml of liquid $^3$He. The microscope is bolted to a multi-pin plug at the bottom of the $^3$He pot which allows an easy exchange. The $^3$He pot has a tenon on its side. It can be clamped at two different heights within the cryostat, i.e., temperature levels (1.5\,K and 4\,K), by means of a cold switch. This switch consists of a guiding rail with a mortise where the tenon can slide into and is clamped by a piston. It can be pressurized from the outside with up to 40\,bar of $^4$He, resulting in a force of 1.6\,kN. One piston with guiding rail is thermally anchored to the 1\,K pot, and one to the bottom of the main $^4$He tank. With this design, a strong clamping force can be applied without putting considerable strain to the sensitive thin rods connecting the parts of the insert vertically. A resisitve heater below the $^3$He pot allows to keep the microscope at elevated temperatures.

A pipe and two empty struts (stainless steel, wall thickness 0.1\,mm) forming a tripod lead upwards from the $^3$He pot to a sorption pump filled with activated carbon. A resistive heater is screwed to the top of the pump for maintaining elevated temperatures above 25\,K. Further upwards, a corrugated hose connects the sorption pump with an external reservoir (volume $0.1\,\mathrm{m}^3$) big enough to hold the entire amount of $^3$He (when gaseous) at a pressure of $8 \cdot 10^4$\,Pa. An external cryopump is permanently attached  so that the insert can be emptied completely from $^3$He.

If the insert is moved down into measurement position, the Au-plated Cu cone of the sorption pump is pressed against a Au-plated Cu counter-cone which is thermally coupled to the 4\,K reservoir via stranded Cu wire. The force ($\approx 350$\,N) is applied to the upper tripod which holds the baffles and is transmitted to the sorption pump through a pressure plate and three pre-loaded springs (5\,kN/m) sitting on top of the sorption pump such that the contact force can be tuned via the $z$ manipulator. Two of the five baffles on top are actively cooled by CuBe springs. They connect the baffle to a copper tube cooled by $^4$He exhaust gas from the main tank.

As temperature sensors, Si diodes are mounted to the upper active baffle and at the pressure plate, and Cernox sensors at the top of the sorption pump, at the top of the $^3$He pot, at the multi-pin plug and within the microscope.

The wiring has been chosen under the aspect of low thermal conductivity and UHV compatibility. Therefore, polyimide-coated Manganin ($\varnothing = 0.1$\,mm) is used for single stranded wires, and stainless steel for coax cables. Thermal anchoring of the wiring is provided at several baffles, along the sorption pump and on the $^3$He pot by winding the wires around small Cu cylinders and fastening them with Kapton foil. Corresponding wires are twisted by hand and surrounded by CuBe mesh for easier handling and shielding.

\subsubsection{Cooling procedure}
We start the description from the following initial state of the cryostat: the LN$_2$ and $^4$He tanks are filled. The direct line between $^4$He tank and 1\,K pot is open and the 1\,K pot is filled with liquid $^4$He at 4.2\,K. The $^3$He is in the gas phase with most of the gas being in the room temperature reservoir. The insert with the microscope is precooled and connected to the 4\,K cold switch. In this situation, the sorption pump is at about 50\,K and holds almost no He.

To start further cooling, the 4\,K switch is opened and the insert is lowered to the height of the 1\,K cold switch which is connected to the 1\,K pot. After closing the cold switch, the direct line to the main tank is closed, and the scroll pump is switched on, pumping on the 1\,K pot. Thereby, the $^3$He pot with the microscope and the 1\,K pot are connected and, hence, cooled simultaneously. During condensation, the pressure in the $^3$He reservoir slowly drops, and so does the temperature in the $^3$He pot until 6800\,Pa or 1.5\,K are reached. Condensation of the full amount of $^3$He from room temperature takes two to three days. 
This large period of time is limited by the heat conductivity of the cold switch (pressed contact) which can be estimated to be of the order of 0.1\,W/K.\cite{berman58} Condensation in the $^3$He pot at a pressure of $8 \cdot 10^4$\,Pa or less can only take place below 3\,K. Therefore, the temperature difference across the cold switch is restricted to about 1\,K and will decrease further upon reduction of the $^3$He pressure. Hence, the heat flow is $\approx 0.1$\,W. This heat flow has to cool the $^3$He being largely at room temperature which requires $\approx 17$\,kJ, such that the cooling time is estimated to be 48\,h, in good agreement with the observation. To a lesser extent, precooling of the gas through the active baffles (Fig.\ \ref{fig:insert}) takes place but is difficult to quantify. After condensation, the cold switch is released, and the insert is moved downwards into measurement position. Once the condensation is finished, the pump of the 1\,K pot can be switched off to reduce vibrations and the 1\,K pot can be refilled with liquid He from the main tank by opening the direct line.

\begin{figure}
\includegraphics{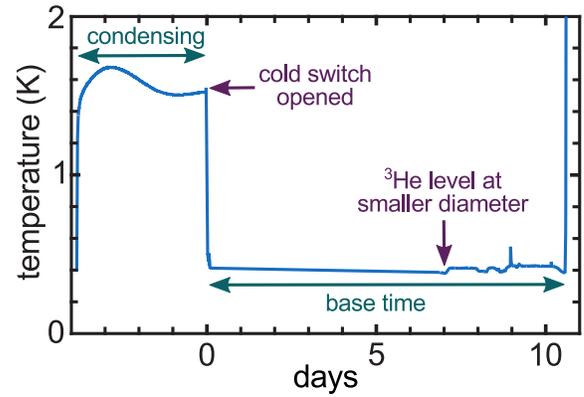}
\caption{\label{fig:hold} Microscope base temperature and hold time. Left part ($<0$ days): initital condensation of $^3$He with the 1\,K cold switch closed. Note that subsequent condensations require only 24 hours. Day 0: end of condensation, release of cold switch and insert moved to measurement position with the sorption pump pressed to the 4\,K counter-cone. After 6.5 days, the temperature changes reproducibly due to the varying cross section of the $^3$He pot and, hence, a varying pumping speed. After 10 days, all $^3$He is evaporated and adsorbed in the sorption pump, such that the temperature rises to a new equilibrium at 9\,K.}
\end{figure}
In measurement position, the cone-shaped bottom of the sorption pump is pressed against a counter-cone thermally coupled to the 4\,K reservoir. Although the thermal conductivity of this junction is comparable to that of the cold switch, the thermal energy of the sorption pump (7\,kJ) is dissipated faster due to the higher temperature gradient. Below 18\,K, the sorption pump starts pumping. It reaches an equilibrium temperature of 9\,K. This way, the $^3$He pot and the attached microscope reach the base temperature of 390\,mK within 3\,h after end of condensation, limited by the time necessary to cool down the sorption pump and achieve full pumping speed. The temperature sequence after the start of condensation is shown in Figure \ref{fig:hold}. After 6.5 days, the temperature rises in a reproducible way to 420\,mK due to a varying cross section of the $^3$He pot (Fig.\ \ref{fig:hold}, bottom right) and remains below 420\,mK as long as there is $^3$He left to evaporate. Temperatures above 420\,mK can be achieved by heating the top of the sorption pump while, at the same time, reducing the pumping speed. The hold time at $390-420$\,mK is more than 10 days. After that, the temperature rises to a new equilibrium at 9\,K with all $^3$He being in the sorption pump.

Regeneration starts by retracting the pump from the cone and closing the 1\,K switch again. The temperature increase of the sorption pump leads to a release of the $^3$He gas. This process can be accelerated by the heaters mounted on the top of the sorption pump. Best condensing performance is achieved when not all $^3$He goes back to the reservoir and therefore condensing is started while regenerating the sorption pump. Then, condensation only takes 24 hours. Above $T_{\mathrm{sorp}} = 55\,\mathrm{K}$ most $^3$He is desorbed.

\subsection{The microscope}
\begin{figure}
\includegraphics{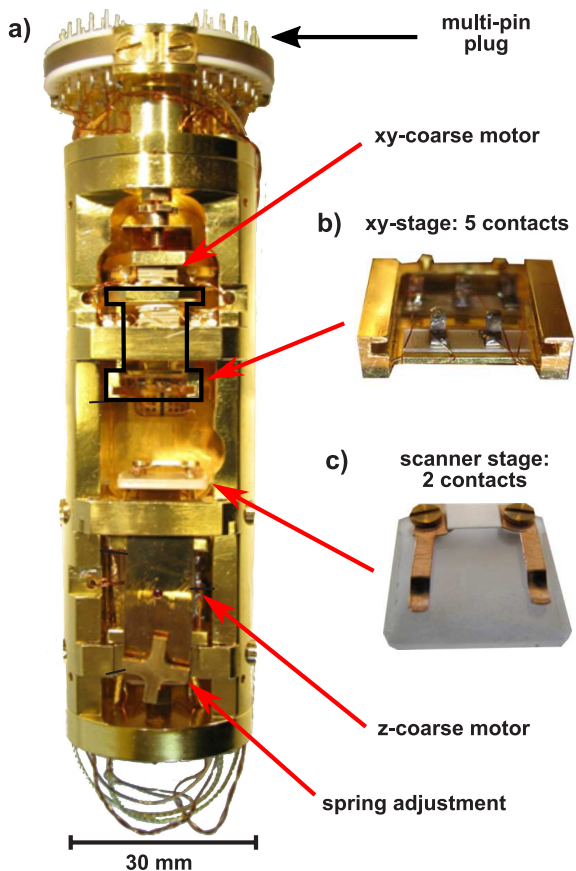}
\caption{\label{fig:microscope} Photographs of the microscope.
(a) Front view (transfer direction). The contour of the $xy$ stage is highlighted by black lines.
(b) $xy$ stage with 5 contacts made from springs glued onto Al$_2$O$_3$ plates.
(c) Scanner stage with two isolated CuBe springs.}
\end{figure}

Attached to the bottom of the $^3$He pot is a home-built scanning probe microscope with an $xy$ positioning stage (Fig.\ \ref{fig:microscope}). With two contacts at the scanner stage and five contacts at the $xy$ stage, it provides several possibilities for tip-sample configurations, including four-point transport measurements of a gated sample with simultaneous scanning tunneling spectroscopy or tuning-fork operated AFM,\cite{giessibl98} if realized by two contacts.\cite{samaddar16}

The microscope body (Fig.\ \ref{fig:microscope}a) is made of phosphorous bronze for good thermal conductance, high mechanical stability, electrical shielding and feasible manufacturing. The resonance frequencies of the two lowest bending modes have been calculated by a finite-element simulation\cite{solidworks} to be 1180\,Hz and 1281\,Hz. The next eigenfrequency, a torsion around the vertical axis, is above 3\,kHz. Thus, all frequencies are above the bandwidth normally used for current detection. The microscope has a multi-pin plug at the top for easy connection to the wiring of the insert, e.g., after modification or repair of the microscope. Two stages are available for tip and sample, both being able to hold carriers of $12 \times 12\,\mathrm{mm}^2$, such that the role of tip and sample can be swapped.

The upper stage offers 5 contacts which are applied to the carrier from below through CuBe springs (Fig.\ \ref{fig:microscope}b). It sits on an $xy$ table which is clamped from either side by two times three piezo stacks mounted horizontally. Pressure to this clamping is applied from the top by a CuBe spring. Its tension can be adjusted {\it in situ} by turning a specially designed screw with a wobble stick. Four shear-mode piezos in each stack realize a stick-slip motion in $x$ and $y$ direction, respectively, and one more piezo plate can wobble vertically in order to excite the tuning fork of the AFM. The lateral displacement of the $xy$ table is up to $2 \times 2\,\mathrm{mm}^2$. A sample inserted into this stage can be inspected optically from below under an angle of about $30^\circ$ using a long-distance microscope with a resolution of $5-10\,\mu\mathrm{m}$. From another flange at the same polar angle, material can be evaporated onto the sample into the cold microscope or, in the same line of sight, onto a sample placed on a stage positioned inside the thermal shielding of the He main tank at a temperature of 30\,K. The latter stage can be used if the microscope should be protected from the deposit.

The other stage of the microscope (Fig.\ \ref{fig:microscope}c) is made out of sapphire and rests on a piezo tube scanner providing a scan range of about $600 \times 600\,\mathrm{nm}^2$ at low temperatures. The lowest bending mode exhibits a resonance frequency of 1.8\,kHz when a sample is inserted. Two contacts are provided by separate springs on the left and right. The sapphire stage is isolated and shielded from the scanner voltage by a grounded layer of conductive epoxy glue on its backside. The scanner is glued into a sapphire prism which is held from three sides by two piezo stacks each, forming a Pan-style stepper motor.\cite{pan92} The pressure between sapphire prism and piezo stacks is applied by a CuBe spring which can be adjusted {\it in situ} by turning a screw with a wobble stick. The two contact springs on the scanner stage and two at the $xy$ stage are connected to coaxial wires. Two more contacts of the $xy$ stage are twisted pair and one is only single stranded due to limited space in the microscope body and may be used as a less sensitive gate contacts.

\subsection{Noise considerations}
The main goal of the wiring is to provide a uniquely defined potential to every point of the system which is electronically relevant. Therefore, multiple grounding at slightly different potentials should be avoided, which can lead to ground loops being susceptible to time dependent magnetic fields. The top flange of the cryostat which hosts all feedthroughs for the insert serves as a ground reference point. It is grounded by an 8\,mm drain wire taken from the rack which contains the electronics for measurements. The ion getter pumps are operated using transformers. Rubber pads isolate the stainless steel frame from the floor, and electrically insulating KF shells break the electrical connection to the pipes of the He recovery. The only other connections to ground are given by the electronics of the Bayard-Alpert gauges which is tested to be not critical.

All wires carrying current or low-bias signals are low-noise micro BNC coax cables. Commercial current-to-voltage converters\cite{femto} are directly connected to the feedthroughs. Bias voltages are transmitted to the insert via differential amplifiers. On all feedthroughs, well grounded filter boxes are mounted which contain $RC$ low-pass filters of different cut-off frequencies to protect the inner part of the UHV chamber from radio frequency noise.

\section{Performance of the instrument}
For the quality of spectroscopic measurements, two aspects are crucial: the spatial resolution and the energy resolution. The primarily desired local density of states (LDOS) can be measured by recording the differential conductivity $dI/dV$. This signal depends critically on the stability of the tunneling current which depends exponentially on the distance. A variation of 1\,pm leads to a 2-3\% change of the current for typical tunneling barriers. Therefore, in order to reach a sufficiently stable signal to noise ratio for spectroscopy, atomic resolution is not enough.

\subsection{Spatial resolution and drift}
\begin{figure}
\includegraphics{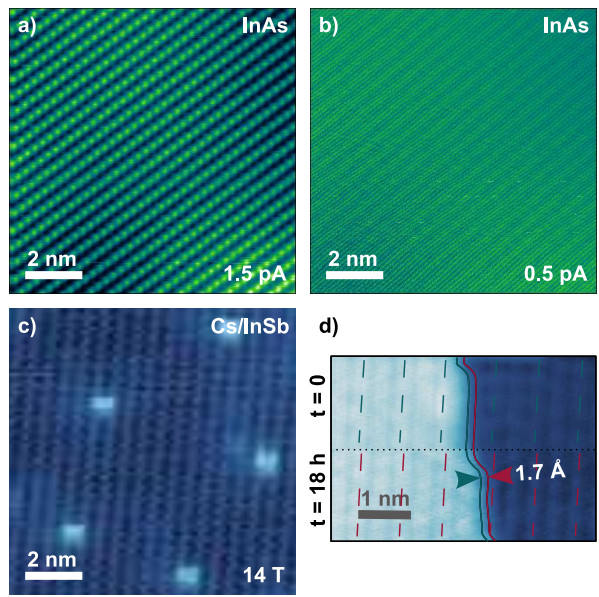}
\caption{\label{fig:xyres} Low-current cut-off, $B$ field stability and thermal drift. Unfiltered topographic images of single crystal surfaces with atomic resolution.
(a) InAs(110), $I = 1.5$\,pA, $V = 500$\,mV, time per line $t_\mathrm{line} = 1$\,s, $10 \times 10\,\mathrm{nm}^2$, $400 \times 400$\,px, $B = 0$\,T, $T = 410$\,mK.
(b) InAs(110), $I = 0.5$\,pA, $V = 500$\,mV, $t_\mathrm{line} = 1$\,s, $10 \times 10\,\mathrm{nm}^2$, $400 \times 400$\,px, $B = 0$\,T, $T = 390$\,mK.
(c) Cs/InSb(110), $10 \times 10\,\mathrm{nm}^2$ cut-out of a $50 \times 25\,\mathrm{nm}^2$ image. $I = 30$\,pA, $V = 300$\,mV, $t_\mathrm{line} = 1$\,s, $400 \times 200$\,px, $B = 14$\,T, $T = 9.2$\,K. Bright dots are Cs atoms; vertical lines: atomic rows.
(d) Sections of two $3.5 \times 5\,\mathrm{nm}^2$ images on InSb(110) recorded at the same area but with 18\,h delay. The solid green (red) line marks the step edge extracted from the first (second) image, the dotted lines the respective atomic rows. The lateral shift is 1.7\,{\AA}. $V = 300$\,mV, $I = 30$\,pA, $t_\mathrm{line} = 5$\,s, $10 \times 10\,\mathrm{nm}^2$, $400 \times 400$\,px, $B = 0$\,T, $T = 434-390$\,mK.}
\end{figure}
Figure \ref{fig:xyres} shows topographic constant-current images of InAs(110) and InSb(110) single crystal surfaces, obtained by cleaving in UHV, which are recorded down to $T=390$\,mK. At a current of $I = 1.5$\,pA, the atomic contrast of the As atoms is clearly visible (Fig.\ \ref{fig:xyres}a). After reducing the tunneling current further to 500\,fA (Fig.\ \ref{fig:xyres}b), atomic rows are still visible (with a superimposed substructure due to residual 50\,Hz noise originating from data communication wiring between scan controller and high-voltage amplifier) but the atomic structure within the rows disappears. Thus, we can probe atomic contrast down to about 0.5\,pA. At increasing magnetic fields, the coupling between microscope insert and He Tank increases due to eddy currents. This initially leads to a better image quality at fields up to 0.5\,T, where external vibrations of the insert are damped by the higher mass of the main tank. At higher field, the coupling is so strong that the effect of the second damping stage is partly bypassed, and vibrations of the cryostat are transmitted to the microscope, reducing the image quality (Fig.\ \ref{fig:xyres}c).

The stability is tested while scanning the same area on an InSb(110) surface for 24 consecutive hours in alternating up and down scans with each image taking one hour. Kinks in an atomic step edge allow the three-dimensional determination of drift. Scanning was commenced 4 days after cool down to 400\,mK in order to exclude thermal drift due to a continuous temperature reduction during consecutive images. The piezo creep was allowed to decay by scanning for 6\,h before taking the first image of Figure \ref{fig:xyres}d. During the following 18\,h, images were acquired without further parameter change. The last image of this sequence is shown in the lower part of Figure \ref{fig:xyres}d. Comparing these two images, a lateral drift of 1.7\,{\AA} and a vertical drift of about 10\,pm has been determined. The speed and direction of the drift varied such that differences between consecutive up-up or down-down scans could be up to 0.3\,{\AA}/h in $xy$ and 0.15\,{\AA}/h in $z$ direction. With this stability, the tip can be held above atomic features for several hours without activation of the feedback. Given the fact that no special precautions to minimize thermally induced drifts are implemented within the microscope, this is a quite satisfactory result. However, as usual, the drift values highly depend on the thermal and electric history, especially of the scanner. In particular, frequent changes of scan area and scanning speed will significantly worsen the stability.

\subsection{Vertical noise}
The most important spatial requirement is the stability of the tip-sample separation. As always when dealing with noise, the bandwidth at which it is measured is crucial for its value. Moreover, when recording images with feedback, noise can show up in two channels, topography and current, and their relative weight is governed by the feedback parameters. In general, noise figures of each channel can be tuned small by the settings of an active feedback.

Therefore, we employed a method which allows inter-system comparison by eliminating the feedback and giving the bandwidth. On an InAs(110) surface after imaging with atomic resolution, $I(z)$ curves are acquired in order to make sure that the tip is stable and provides normal vacuum-tunneling conditions. These curves provide a unique relation between distance and current such that, determined by the bandwidth of the preamplifier, the $z$ stability can be deduced from the current recorded over time with the feedback switched off, provided there are no other sources of current noise. Hence, we get a safe upper bound for the $z$ noise. The current time trace is converted into a $z$ time trace and Fourier transformed revealing the spectral $z$ noise density. This is the relevant noise figure when it comes to spectroscopy experiments.

\begin{figure}
\includegraphics{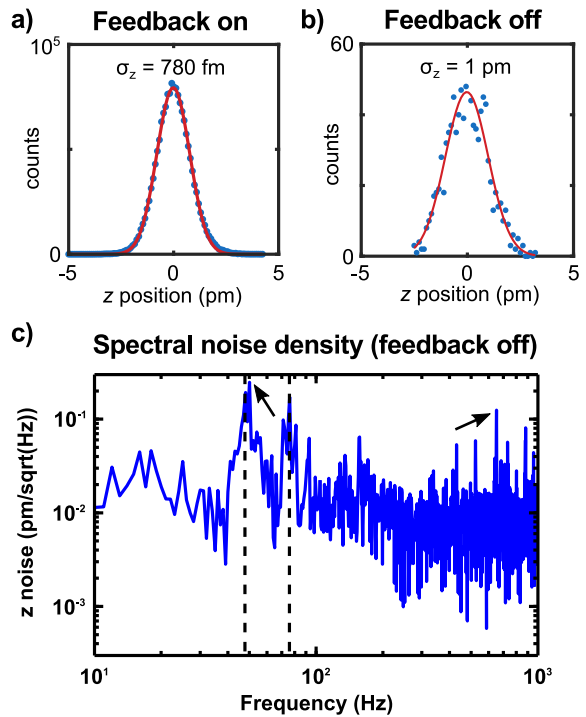}
\caption{\label{fig:noise_spec} Vertical noise and spectroscopic stability. (a) Histogram of recorded $z$ positions while scanning above a single point with activated feedback (blue dots) and Gaussian fit (solid red line) with standard deviation $\sigma_z$ ($400 \times 400$\,px, 0.5\,s/line, $V = 500$\,mV, $I_\mathrm{stab} = 150$\,pA). (b) Histogram of the tunneling current (1000 points per 0.5\,s, $V = 500$\,mV, $I_\mathrm{stab} = 150$\,pA), acquired above one location for 1\,s without feedback and converted into height information according to $I(z)$ curves, $I(z)=I_0 \exp^{-\kappa z}$ with $\kappa = 13.7\,\mathrm{nm}^{-1}$. Data and fit as in (a). (c) Spectral noise density of the measurement shown in (b). Dashed lines (black arrows) mark peaks due to mechanical (electrical) noise. Bandwidth: 700\,Hz.}
\end{figure}

Figure \ref{fig:noise_spec}c shows the spectral $z$ noise density using an exponential fit to the $I(z)$ curve: $I(z)=I_0 \exp^{-\kappa z}$ with $\kappa = 13.7\,\mathrm{nm}^{-1}$. The noise floor is mainly due to electronic white noise. The residual peaks at 50 and 620\,Hz are of electronic origin, hence, also present when the tip is retracted. The broader peak structures around 45\,Hz and 75\,Hz are the dominant mechanical perturbations arising from gas oscillations in the $^4$He exhaust lines. The frequency and amplitude of these gas oscillations can be tuned by tuning the valves of the exhaust lines.

Hence, we provide a $z$ noise of Gaussian width $\sigma_z = 1$\,pm at a bandwidth of 700\,Hz. Other systems, which provide STM measurements under UHV conditions at $T<1$\,K with optical access in order to place the tip with $10\,\mu$m precision, have all been at or above 2\,pm.\cite{wiebe04,song10,zhang11,misra13,allwoerden16} Consequently, we conjecture that the compact design -- made possible by the all-UHV manufacturing of the cryostat -- is indeed beneficial for the noise performance.

\begin{figure}
\includegraphics{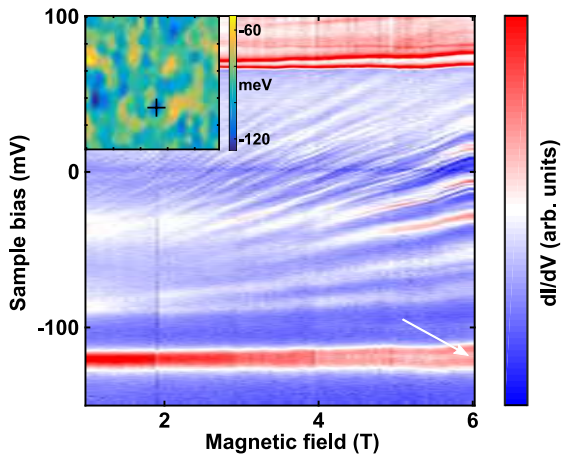}
\caption{\label{fig:Landau_fan} Spectroscopic stability. Main image: differential conductivity $dI/dV(V,B)$ on a Cs/InSb(110) single crystal taken at the position marked by the cross in the inset. The two lowest Landau levels are marked by a white arrow. Inset: potential map on an $250 \times 250\,\mathrm{nm}^2$ area showing the average energy of the two lowest Landau levels taken from a $dI/dV(x,y,V,B=6\,\mathrm{T})$ map.\cite{bindel16} Stabilization parameters: $V=50$\,mV, $I=0.1$\,nA. Bias modulation: $V_\mathrm{ac}=0.75$\,mV.}
\end{figure}

The stability of the instrument for scanning tunneling spectroscopy is demonstrated by STS maps in magnetic fields up to 14\,T. They were used to study the local variation of the Rashba parameter of a two-dimensional electron system on an InSb(110) single crystal in the quantum Hall regime,\cite{bindel16} and the nodal structure of Landau level wave functions.\cite{bindel17} Figure \ref{fig:Landau_fan} shows a Landau fan (differential conductivity as a function of $B$ field and sample voltage) at a potential minimum. The potential landscape has been acquired by evaluating the average energy of the lowest spin-split Landau level from a spectroscopy map $dI/dV(x,y,V)$ at $B=6$\,T. Subsequently, the Landau fan was recorded by in turns measuring $dI/dV(V)$ at a chosen position and ramping the $B$ field. After fitting with Lorentzian peaks, spin splittings were extracted with a precision of better than 0.5\,meV for 65\% of the data points.

\subsection{Electronic temperature}
The energy resolution of an STS experiment is not only given by noise but also by the sharpness of the Fermi-Dirac distribution of the probing tip. Since the electron system may decouple from the bulk crystal due to suppression of electron-phonon interactions and due to the increasing ratio of electronic vs.\ phononic thermal conductivity at low temperatures, it is necessary to independently determine the electronic temperature.

\begin{figure*}
\includegraphics{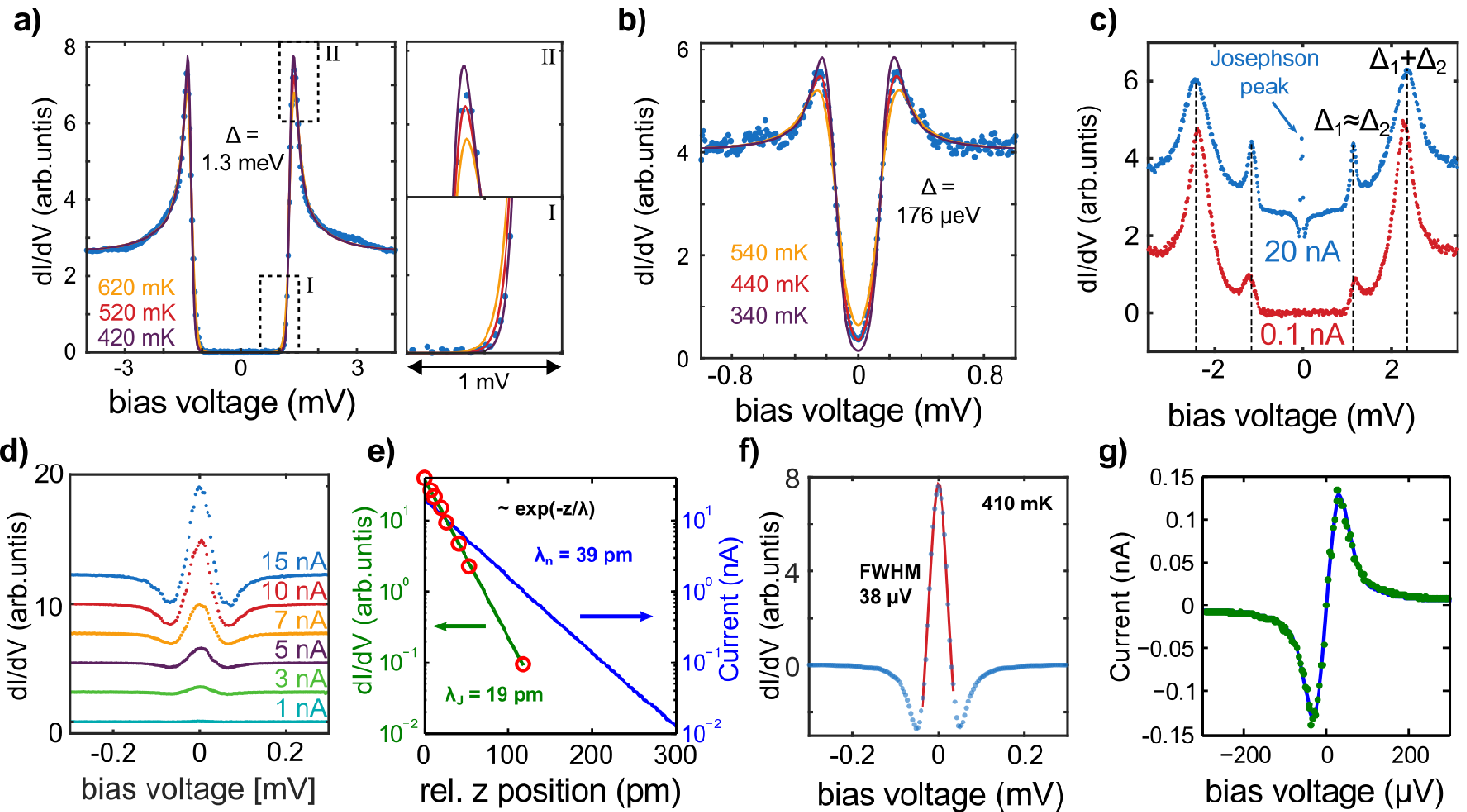}
\caption{\label{fig:supercond} Electronic temperature and energy resolution.
(a) Differential conductivity $dI/dV$ of a normal conducting tip on Pb/W(110) (blue) obtained at $T = 393$\,mK and $B=0$\,T. Fit (red) using the lock-in amplitude of $V_\mathrm{mod} = 35\,\mu V$ and an electronic temperature of the tip $T_\mathrm{el}^\mathrm{tip} = (520 \pm 20)\,\mathrm{mK}$, graphs in orange, purple for different $T_\mathrm{el}^\mathrm{tip}$ as marked. $V_\mathrm{stab} = 4$\,mV, $I_\mathrm{stab} = 100$\,pA. Insets I and II are zoom-ins for judging the fit quality.
(b) Nb tip on W(110) at 393\,mK. Fit (red) for $T_\mathrm{el}^\mathrm{sample} = (440 \pm 9)\,\mathrm{mK}$ and $V_\mathrm{mod} = 35\,\mu V$, graphs in orange, purple for different $T_\mathrm{el}^\mathrm{sample}$ as marked.
(c-g) Josephson tunneling across an S-S tunnel junction ($V_\mathrm{stab} = 4$\,mV, $V_\mathrm{mod} = 9\,\mu$V, $T = 410$\,mK).
(c) Differential conductivity for two stabilization currents (0.1\,nA and 20\,nA). Peaks labeled by $\Delta_{1,2}$ arise from multiple Andreev reflections, the zero-bias peak at 20\,nA is due to Josephson tunneling.
(d) Development of the Josephson peak intensity for decreasing tip-sample distances (higher stabilization current, as indicated).
(e) Height of the Josephson peak (red circles) vs.\ tip-sample distance with exponential fit $\propto e^{-z/\lambda}$ (green, $\lambda = 19$\,pm) and tunneling current at $V = 5$\,mV vs.\ distance (blue, $\lambda = 39$\,pm).
(f) Josephson peak (blue) for $I_\mathrm{stab} = 20$\,nA with Lorentzian fit (red).
(g) $I(V)$ curve of the Josephson current (green) fitted by the $P(E)$ model (blue).\cite{ast16}}
\end{figure*}

To this end, we employed tunneling spectroscopy with normal and superconductors in different tip-sample configurations. A W(110) single crystal has been annealed in UHV, and 20 monolayers of lead (bulk energy gap: $\Delta = 1.3\,\mathrm{meV}, T_c = 7.19$\,K) were evaporated onto the crystal. Figure \ref{fig:supercond}a shows STS data of the local density of states (LDOS) obtained at 393\,mK with a Nb tip. A single gap of 1.3\,meV close to the bulk value of Pb is found, implying a superconducting sample probed by a normal-conducting tip. The absence of superconductivity of the tip might be related to the sensitivity of superconducting properties on details of the tip apex such as composition, shape and structure.\cite{pan98} Fitting the Bardeen-Cooper-Schrieffer density of states, $\rho_\mathrm{SC} = \rho_0 \cdot \frac{|E|}{\sqrt{E^2-\Delta^2}}$,\cite{bardeen57} convolved with the Fermi-Dirac function $f(E,T_\mathrm{el}^\mathrm{tip})$ and taking the lock-in amplitude ($V_\mathrm{mod} = 35\,\mu$V) into account\cite{pan98,wiebe04} yields $T_\mathrm{el}^\mathrm{tip} = 520 \pm 20$\,mK for the sample on the piezo scanner and the tip on the $xy$ stage.

To address the electronic temperature of the sample, the tip (mounted on the $xy$ stage) is chosen to be superconducting, i.e., the Nb tip has been prepared by voltage pulsing until it shows a superconducting gap on a W(110) single crystalline sample. The gap is considerably smaller than for bulk Nb ($\Delta = 1.4\,\mathrm{meV}, T_c = 9.26$\,K), likely due to the proximity effect,\cite{pan98} and therefore the differential conductivity around zero bias is particularly sensitive to the electronic temperature of the sample, which, in this case, is found to be $T_\mathrm{el}^\mathrm{sample} = 440 \pm 10$\,mK (Fig.\ \ref{fig:supercond}b).

Thus, we concede that the electronic temperature is a factor of $1.1-1.3$ larger than that of the phonon bath which is probed by the temperature sensors. Other $^3$He STM systems also achieve $T_\mathrm{el}/T_\mathrm{ph} \approx 1$,\cite{wiebe04,kamlapure13} while for STMs in pulse tube systems and dilution refrigerators reaching $T_\mathrm{ph}<100$\,mK, this factor varies from 2.4 to 12\cite{kambara07,assig13,misra13,pelliccione13,singh13,roychowdhury14,galvis15} with the lowest electronic temperature reported in such systems being 38\,mK.\cite{assig13}

\subsection{Energy resolution}
To estimate the energy resolution without being dominated by the broadening effect of $f(E,T)$ by the Fermi-Dirac distribution, tunneling between superconductors is investigated. A superconductor-superconductor junction is prepared by dipping the tip into the Pb film, and a double gap is observed (Fig.\ \ref{fig:supercond}c), as expected from tunneling involving multiple Andreev reflections.\cite{ternes06} Reducing the tunneling barrier by increasing the stabilization current (Fig.\ \ref{fig:supercond}d) leads to the development of a peak at zero bias associated with Josephson tunneling of Cooper pairs.\cite{naaman01} This is corroborated by the distance dependence of the zero-bias peak height in comparison with that of the tunneling current of electron-like quasiparticles away from the superconducting gap (Fig.\ \ref{fig:supercond}e). Both show an exponential decay, but with a decay length of 39\,pm for the quasiparticles and 19\,pm for the Cooper pairs. The factor of two in the decay length is expected: while the normal state tunneling current $I_n$ is proportional to the conductivity of the tunnel barrier $I_n \propto \sigma_n \propto e^{-z/\lambda}$, the Josephson current $I_\mathrm{J}$ is proportional to the square of the Josephson energy $E_\mathrm{J}$,\cite{ingold94,jaeck16} which in turn depends on the critical current $I_\mathrm{C}$ and the normal-state conductivity of the barrier: $E_\mathrm{J} \propto I_\mathrm{C} \propto \sigma_n$.\cite{ambegaokar63} So we have $I_\mathrm{J} \propto E_\mathrm{J}^2 \propto \sigma_n^2 \propto e^{-2z/\lambda}$.

Fitting a Lorentzian curve to the Josephson peak yields a FWHM of $38\,\mu\mathrm{V}$ (Fig.\ \ref{fig:supercond}f), significantly smaller than the temperature broadening of the Fermi-Dirac function ($\approx 3.5 k_\mathrm{B}T$). This is not surprising since Cooper pairs condensed into a BCS state do not obey Fermi-Dirac statistics. Instead, they are condensed to a single energy such that broadening is not determined by the distribution function of the electrodes. A simple possibility to explain the finite width of the Josephson peak is a remaining voltage noise between tip and sample. In turn, we can take the $38\,\mu\mathrm{V}$ as an upper bound of this noise. Indeed, we used the width of the Josephson peak to reduce voltage noise by adequate $RC$ filtering which, however, has not been further improved beyond the $38\,\mu\mathrm{V}$ limit.

Another -- intrinsic -- possibility to explain the remaining width of this peak is the interaction of the tunneling Cooper pairs with the electromagnetic surroundings.\cite{jaeck16} The probability of a tunneling electron to exchange photons with the environment mediated by the capacitor formed by tip and sample, is therefore described by a probability function called $P(E)$. Figure \ref{fig:supercond}g shows a fit of an $I(V)$ curve stabilized at 4\,mV and 20\,nA, according to the method described in Ref.\ [\onlinecite{ast16}]. The $P(E)$ function depends on the temperature $T$, the capacitance of the tunnel junction $C_\mathrm{J}$ and the environmental impedance $Z(\omega)$. In our case, we used a modified infinite transmission line impedance to model the environment in the STM\cite{jaeck16} with the principal resonance frequency ($\omega_0 = 70\,\mu$eV), a damping factor ($\alpha = 0.9$) and a zero-frequency resistance being equal to the vacuum resistance ($R_\mathrm{DC}=377\,\Omega$). These parameters can in general be obtained from resonances in the $I(V)$ characteristic and have been shown to vary only slightly from tip to tip.\cite{jaeck15} In this particular curve, resonances were absent, so typical values of $\omega_0$, $\alpha$ and $R_\mathrm{DC}$ have been chosen.

The temperature ($T=410$\,mK) is obtained from the microscope Cernox sensor. The only two fit parameters are the junction capacitance $C_\mathrm{J} = 7.8 \pm 0.2$\,fF and the critical current $I_\mathrm{C} = 11.2 \pm 0.1$\,nA. The found capacitance is in line with calculations for conical tips ($3-21$\,fF for wire diameters between 0.25\,mm and 1\,mm, opening angle: $60^\circ$).\cite{ast16}
The critical current agrees well with the expectation of 10.6\,nA calculated from the Ambegaokar-Baratoff formula $I_\mathrm{C} = \frac{\pi}{2}\frac{\Delta}{R_\mathrm{N}}$ with the normal-state resistance of the junction given by the stabilization parameters ($\frac{4\,\mathrm{mV}}{20\,\mathrm{nA}}=R_\mathrm{N} = 200\,\mathrm{k}\Omega$).\cite{ambegaokar63} The excellent agreement of the data and the fit show that the noise figures of the instrument are close to the physical limit given at the base temperature.

\subsection{Transport measurements}

An extraordinary feature of our UHV-STM setup is the possibility to probe magnetotransport simultaneously with STM or AFM measurements. This allows in particular to drive the system into a desired transport regime prior to probing the responsible electron wave functions, e.g., at the Fermi level. To this end, we demonstrate the excellent quality of the transport data within the STM.

\subsubsection{Shubnikov-de Haas oscillations of a two-dimensional electron system in GaAs}
\begin{figure}
\includegraphics{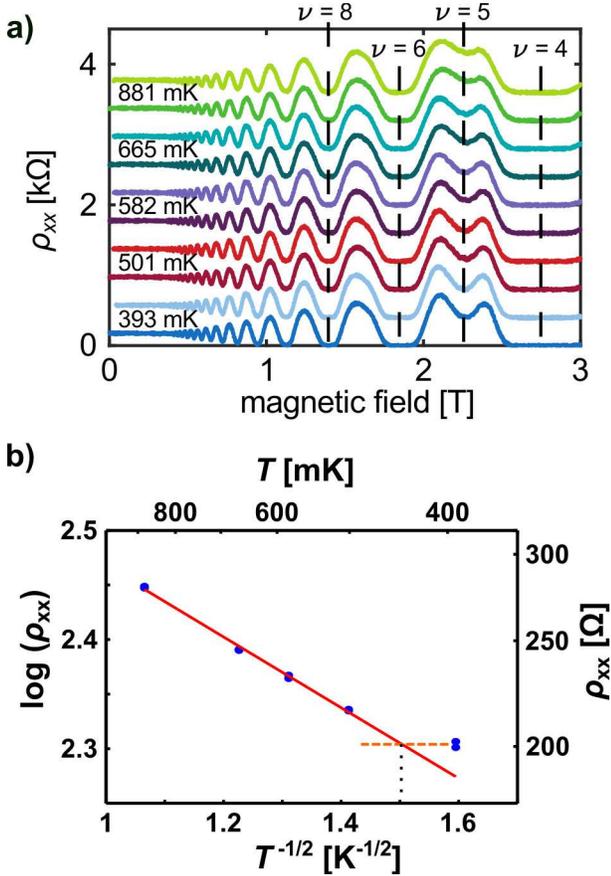}
\caption{\label{fig:SdH} Shubnikov-de Haas oscillations of a GaAs/AlGaAs 2DES probed within the STM. (a) Longitudinal resistance $\rho_{xx}$ as a function of perpendicular magnetic field at various temperatures as indicated, recorded by a Cernox sensor within the microscope at $B=0$\,T. Curves are offset for clarity. Filling factors $\nu$ are deduced from $1/B$ plots. (b) The minimum resistance at filling factor $\nu = 5$ over temperature in an Arrhenius-like plot with red fit line using $T_0 = 557$\,mK. The deviation of $\rho_{xx}$ at $T=393$\,mK is interpreted as saturation of the electronic temperature $T_\mathrm{el}^\mathrm{sample}$. Projection to the fitting line gives $T_\mathrm{el}^\mathrm{sample} = 440$\,mK.}
\end{figure}

We made use of the temperature-dependent Shubnikov-de Haas (SdH) oscillations in a two-dimensional electron system (2DES).\cite{matthews05} A GaAs/AlGaAs heterostructure was contacted using four contacts of the $xy$ stage. SdH oscillations were recorded within the STM in UHV and at measurement position as longitudinal resistance $\rho_{xx}$ vs.\ magnetic field for different temperatures (Fig.\ \ref{fig:SdH}).
At $B=2.2$\,T, the spin splitting at filling factor $\nu=5$ can be resolved (Fig.\ \ref{fig:SdH}a). The resistance at the minima is given by thermal excitations of electrons to other edge channels or, for lower temperatures, by variable-range hopping (VRH).\cite{efros75} Above 500\,mK, the resistance follows the expected model for VRH according to $\rho_{xx} \propto \exp(-\sqrt{T_0/T})$ (Fig.\ \ref{fig:SdH}b). Only the resistivity at 393\,mK deviates significantly from the fit which indicates that the temperature of the electron system does not follow the temperature of the phonons anymore. Reading out the resistance at this temperature and projecting it onto the fitting curve, we find an estimate of the electronic temperature, $T_\mathrm{el}^\mathrm{sample} = 440 \pm 20$\,mK, which is in excellent agreement with $T_\mathrm{el}^\mathrm{sample}$ found by probing in STM with a superconducting tip (Fig.\ \ref{fig:supercond}b).

\subsubsection{Landau fan of graphene on boron nitride}
\begin{figure}
\includegraphics{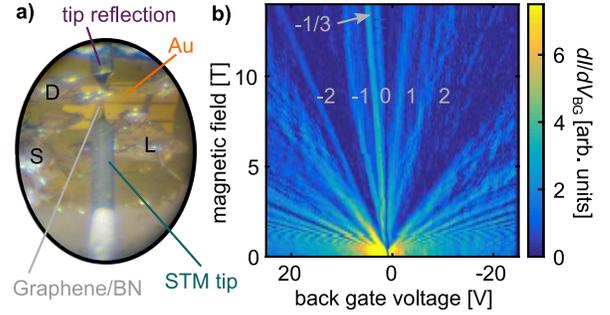}
\caption{\label{fig:graphene} Landau fan of graphene in the quantum Hall regime. (a) Optical view through the long-distance microscope onto the sample inside the STM at $T \approx 25$\,K while positioning the tip on the landing area. Gold pads for source (S), drain (D) and tip landing area (L) are contacted by conductive silver and are used to navigate on the sample in order to find the graphene flake ($10 \times 20\,\mu\mathrm{m}^2$, not visible, position between gold pads as indicated). (b) Landau fan acquired in a two-terminal measurement within the $xy$ stage of the microscope by recording the transconductance $dI/dV_\mathrm{BG}$ across the sample at oscillating bias voltage $V_\mathrm{AC,SD} = 100\,\mu$V, $f=654$\,Hz, $T=1.6$\,K. The sample is inside the STM in measurement position, numbers indicate filling factors.}
\end{figure}
The ability to find microstructured samples with an STM tip and to perform transport measurements on these samples {\it in-situ} is demonstrated by exfoliation of graphene and boron nitride in order to form a stack on the surface of a SiO$_2$/Si substrate using a dry transfer method.\cite{kretinin14} A mesh of gold pads has been evaporated onto the sample, providing a landing area for the tip. Optical navigation is possible by moving the $xy$ stage under inspection with a long-distance microscope through which the tip image and the tip reflection can be inspected down to the $10\,\mu$m range (Fig.\ \ref{fig:graphene}a). The sample temperature stays below 25\,K during this procedure.

For magneto-transport measurements, an oscillating source-drain voltage of $V_\mathrm{AC,SD} = 100\,\mu$V is applied to the graphene sample while the backgate voltage $V_\mathrm{BG}$ is ramped between -25\,V and +25\,V and the magnetic field from 0 to 14\,T. The transconductance $dI/dV_\mathrm{BG}$ is displayed after numerical derivation of the current (Fig.\ \ref{fig:graphene}b). Conductance minima are found at integer filling factors including those that correspond to symmetry-broken states ($\nu = -1, 0, +1$) as well as one minimum indicating a gap at $\nu = -1/3$. Hence, samples with novel types of electron phases become accessible in a controlled fashion by our combined STM / magneto-transport setup.

\subsection{Atomic force microscopy}

\begin{figure}
\includegraphics{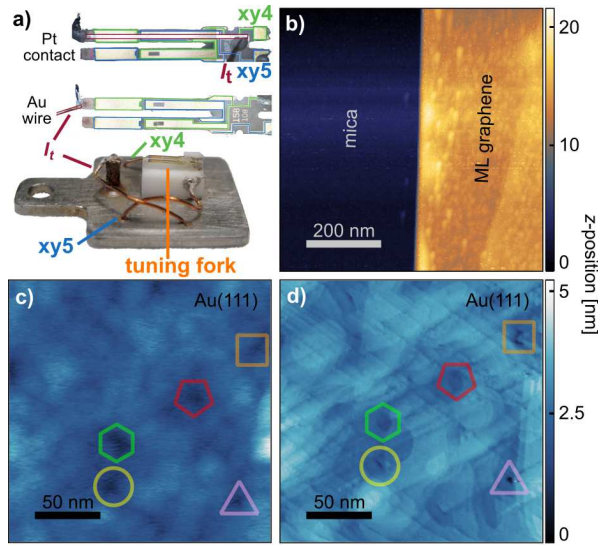}
\caption{\label{fig:afm} AFM with qPlus sensor.
(a) Top: Photograph of the tuning fork with two different ways of contacting the tunneling tip attached to the front of the qPlus sensor. Vapour deposited Platinum contact (red) and gold wire at the front (red). The contact for the tunneling current $I_\mathrm{t}$ and the tuning fork contacts xy4 and xy5 are marked, and the two electrodes of the tuning fork are indicated in green and blue.
Bottom: Photograph of a qPlus sensor mounted on a sample holder with a gold wire for contacting the tunneling tip. 
(b) Edge of an electrically contacted multi-layer graphene sheet on mica ($V_\mathrm{bias} = 0.88$\,V, frequency shift: $\Delta f = -0.35$\,Hz, tip oscillation: $A_\mathrm{tip} = 1.4$\,nm).
(c) Au(111) surface scanned in AFM mode ($V_\mathrm{bias} = 0.5$\,V, $\Delta f = -0.7$\,Hz, $A_\mathrm{tip} = 5$\,nm) and
(d) subsequently in STM mode ($I = 0.5$\,nA, $V_\mathrm{bias} = 1$\,V).
Corresponding features are marked in both images. Images taken under ambient conditions.}
\end{figure}

While extended flakes with transport contacts can still be investigated in pure STM mode, nanostructured samples or investigations at the edge of flakes require insulating areas close to the probed ones and, thus, a different feedback mechanism for distance control. Such a possibility is provided by AFM. With five contacts on the $xy$ stage, it is possible to operate an AFM mode using a tuning fork in different configurations, e.g., as a qPlus sensor\cite{giessibl98} with two contacts for reading out the oscillation via the tuning fork electrodes, two contacts for applying an oscillation excitation to a small piezo stack below the tuning fork, and one contact for detecting the tunnel current through a separate gold wire attached to the tip. However, best results in terms of crosstalk and mechanical quality factor $Q$ have been achieved by using the excitation piezos below the stepper piezos of the $xy$ table (Fig.\ \ref{fig:microscope}a) for excitation and a Pt electrode evaporated onto an insulating film on the tuning fork (top of Fig.\ \ref{fig:afm}a) for detecting the tunneling current. In this configuration, the resonance frequency is 21.9\,kHz with $Q=25000$ in UHV at 400\,mK.

The navigation on insulating samples is demonstrated for an electrically contacted graphene sheet on mica (Fig.\ \ref{fig:afm}b). Imaging the same area on Au(111) in AFM mode and subsequently in STM mode (Fig.\ \ref{fig:afm}c,d) shows that, even during rough imaging under ambient conditions, the main features agree and AFM can be used for orientation keeping the tip intact for STM.

\section{Summary}
To conclude, we have successfully fabricated the first scanning probe microscope allowing for STM, AFM and magneto-transport at the same stage in an all-UHV cryostat operating at temperatures as low as 400\,mK and a magnetic field as high as 14\,T.
We obtain a $z$ noise of Gaussian width $\sigma=1$\,pm in open feedback loop at a bandwidth of 700\,Hz, imaging exhibiting atomic features down to 500\,fA, a thermal drift of 0.3\,{\AA}/h, an electronic temperature of $400-500$\,mK, a voltage noise in the tunnel junction below $40\,\mu$eV and tip navigation within the STM at $\approx 10\,\mu$m precision. magneto-transport abilities in the STM showing, e.g., $\nu=-1/3$ features for graphene on BN have been demonstrated as well as the ability to use the AFM mode for positioning the tip prior to STM measurements.
This novel combination of STM, AFM and transport measurements will allow the \textit{in-situ} spatial imaging of microstructured samples on insulating substrates while simultaneously performing transport experiments. We expect that it will contribute significantly to the bridging between the surface science and the quantum transport community.

% If in two-column mode, this environment will change to single-column format so that long equations can be displayed. 
% Use only when necessary.
%\begin{widetext}
%$$\mbox{put long equation here}$$
%\end{widetext}

% Figures should be put into the text as floats. 
% Use the graphics or graphicx packages (distributed with LaTeX2e).
% See the LaTeX Graphics Companion by Michel Goosens, Sebastian Rahtz, and Frank Mittelbach for examples. 
%
% Here is an example of the general form of a figure:
% Fill in the caption in the braces of the \caption{} command. 
% Put the label that you will use with \ref{} command in the braces of the \label{} command.
%
% \begin{figure}
% \includegraphics{}%
% \caption{\label{}}%
% \end{figure}

% Tables may be be put in the text as floats.
% Here is an example of the general form of a table:
% Fill in the caption in the braces of the \caption{} command. Put the label
% that you will use with \ref{} command in the braces of the \label{} command.
% Insert the column specifiers (l, r, c, d, etc.) in the empty braces of the
% \begin{tabular}{} command.
%
% \begin{table}
% \caption{\label{} }
% \begin{tabular}{}
% \end{tabular}
% \end{table}

\begin{acknowledgments}
We would like to thank A.\ Khajetoorians, J.\ Wiebe, M.\ Bode, M.\ A.\ Krzyzowski and M.\  Wolfram for helpful discussions. Funding through the DFG via INST 222/776-1 FUGG is greatfully acknowledged.
\end{acknowledgments}

%\bibliography{mliebman}
\bibliographystyle{apsrev}

\end{document}